\documentclass[aps,pra,twocolumn,longbibliography,groupedaddress,showpacs,floatfix,superscriptaddress]{revtex4-1}
\usepackage{mathtools,amssymb,graphicx,units}
\usepackage[plainpages=false,pdfpagelabels,colorlinks=true,linkcolor=red,urlcolor=blue,citecolor=blue,pdftitle={Title},pdfauthor={},pdfdisplaydoctitle=true,pdfduplex=DuplexFlipLongEdge]{hyperref}
\usepackage{soul} 
\usepackage[dvipsnames,usenames]{color}
\usepackage[normalem]{ulem}
\usepackage{graphicx}
\usepackage{xcolor}
\usepackage{bbold} 
\usepackage{float}
\emergencystretch=\maxdimen
\def\s{\sigma}
\def\ave#1{\langle #1\rangle}

\newcommand{\iv}{\mathbf{i}}
\newcommand{\jv}{\mathbf{j}}

\newcommand{\up}{\uparrow}
\newcommand{\dn}{\downarrow}

\begin{document}

\title{Magnetic impurities in a charge-ordered background}

\author{Sebasti\~ao dos Anjos \surname{Sousa-J\'unior}} 
\email{sebastiaojr@pos.if.ufrj.br}
\affiliation{Instituto de F\'\i sica, Universidade Federal do Rio de Janeiro
Cx.P. 68.528, 21941-972 Rio de Janeiro RJ, Brazil}
\author{Raimundo R. \surname{dos Santos}} 
\affiliation{Instituto de F\'\i sica, Universidade Federal do Rio de Janeiro
Cx.P. 68.528, 21941-972 Rio de Janeiro RJ, Brazil}
\author{Natanael C. Costa}
\affiliation{Instituto de F\'\i sica, Universidade Federal do Rio de Janeiro
Cx.P. 68.528, 21941-972 Rio de Janeiro RJ, Brazil}

\begin{abstract}
We investigate how magnetic impurities may affect a system exhibiting charge-density wave (CDW) in its ground state. We consider a disordered Hubbard-Holstein model with a homogeneous electron-phonon interaction, but with a (randomly chosen) fraction of sites displaying a non-zero Coulomb repulsion, $U$, and perform state-of-the-art finite-temperature quantum Monte Carlo simulations. For a single magnetic impurity, charge-charge correlations hamper the spin-spin ones around the repulsive site, thus requiring a strong enough value of $U$ to create non-negligible antiferromagnetic (AFM) correlations. As the number of magnetic impurities increases, these AFM correlations become deleterious to CDW order and its features. First, the critical temperature is drastically reduced, and seems to vanish around 40$\%$ of impurities (for fixed $U/\lambda=2$), which we correlate with the classical percolation threshold. We also notice that just a small amount of disorder suffices to create a \textit{bad insulating} state, with the suppression of both Peierls and spin gaps, even within the charge-ordered phase. Finally, we have also found that pairing correlations are enhanced at large doping, driven by the competition between CDW and AFM tendencies.
\end{abstract}
\maketitle
\section{Introduction}
\label{sec:intro}

Over the past decades, much interest has been given to unveil the nature and interplay between long-range ordered phases in transition-metal dichalcogenides (TMDs)\,\cite{Manzeli17,Zhu15,Zhu17}. 
A great experimental effort through different scenarios has been invested to understand the occurrence of charge-density wave (CDW) and superconductivity (SC) in these compounds 
-- from gate-induced\,\cite{Li16} and hydrostatic pressure\,\cite{Kusmartseva2009} to chemical doping\,\cite{Wagner08} and substitutional disorder\,\cite{Li17} --, even though the emergence and competition between these phases are still open issues.
Interestingly, the phase diagrams of TMDs~\cite{Kusmartseva2009} resemble those of doped high-temperature cuprate
superconductors, which has raised the possibility of investigating pseudo-gap phenomena in the former to further understand the latter\,\cite{Chatterjee15}.
However, a still open question about TMDs, and more generally about the nature of the charge~\cite{Ugeda2016} and pairing interplay, is how spin-spin correlations may affect this competition.
In other words, how repulsive electron-electron ($e$-$e$) and retarded electron-phonon ($e$-$ph$) interactions affect the ground state and thermodynamic properties of such compounds. 

Within this context, the TMDs provide odd opportunities to investigate this interplay.
For instance, most of the 1T polytypes~\cite{Kusmartseva2009,Joe2014} exhibit flat bands, a feature that renders both $e$-$e$ and $e$-$ph$ interactions non-negligible.
Beyond this case, the effects of spin-spin correlations on CDW and SC phases may be investigated by systematically intercalating magnetic ions, such as Fe atoms, between layers of a given TMD.
As a direct consequence of such procedure, it has been established that a small amount of doping/intercalation is already enough to suppress the CDW order, and to change the SC critical temperature~\cite{Dai1993,Yan_2019}.
In addition, the effects on transport properties have been examined, e.g., for NbSe$_2$ intercalated by Fe ions~\cite{nair2020}, providing evidence that electric current could be used to adjust the magnetic orientation of the spins, which can make this material suitable for spintronic devices.
Other important features, such as the occurrence of the Kondo effect and its relevance to transport properties are still under intense debate~\cite{Iavarone_2009,Iavarone2011,Pervin2020}.

In order to investigate fundamental properties of such a interacting compounds, one should examine the features of simplified effective Hamiltonians.
Within this context, the single-band Hubbard-Holstein model (HHM)~\cite{berger95} takes into account the Coulomb repulsion between electrons, as well as an indirect retarded electronic interaction due to an \textit{e-ph} coupling.
The inclusion of these interactions may lead to electronic instabilities, with the enhancement of strong charge, spin, and/or pairing correlations -- therefore capturing the interplay between CDW, AFM and/or SC phases.
For instance, the ground state of the pure Holstein model on a half-filled square lattice has been extensively scrutinized\,\cite{scalettar89,Vekic92,Vekic93,Hohenadler04,Weber18}, exhibiting a CDW for any \textit{e-ph} interaction\,\cite{ncc20}.
However, this ordered phase is unstable under external parameters, with the enhancement of (conventional) pairing correlations when doping\,\cite{Dee19,Bradley21}, pressure or strain\,\cite{CohenStead19,Araujo22}, nonlinear \textit{e}-\textit{ph} couplings\,\cite{Li16a,Dee20,Paleari21}, Anderson disorder~\cite{Xiao2021} or phonon dispersion~\cite{ncc2018} take place.
The addition of a repulsive Hubbard-like term to the Holstein model leads to similar behavior: the \textit{e-e} interaction suppresses double occupation, destroying the CDW phase, while enhancing AFM or pairing correlations\,\cite{ncc20,Wang20}.

In this work, we examine a case interpolating between the pure Holstein model and the HHM, in the sense that an \textit{e-e} interaction is only considered on a fraction of sites/orbitals -- i.e., when a percentage of sites are treated as \textit{e-e} interacting.
Such ``dilute-to-dense'' crossover may be a rough model for intercalated magnetic impurities on TMD's, with the \textit{e-e} interacting sites playing the role of an intercalated magnetic site.
In particular, we investigate the stability of the well-known staggered CDW phase on the half-filled square lattice as the number of impurity sites increases.
To this end, we perform unbiased quantum Monte Carlo (QMC) simulations aiming to analyze three main aspects:
the behavior of charge and spin correlations [i] in the dilute regime -- one or two magnetic impurities -- and [ii] in the dense regime, as well as [iii] the behavior of thermodynamic quantities.
Within such analyses, we expect to provide further insights about the nature of the charge ordered phase in the HHM.
The paper is organized as follows:
the model and methodology are outlined in the next Section, while our results are presented in Sections\,\ref{sec:dilute} and \ref{sec:dense}.
Our final remarks are given in Section\,\ref{sec:concl}.

\section{Model and methods}
\label{sec:model}

The Hubbard-Holstein model describes electrons on a lattice interacting with each other through both a direct on-site Coulomb repulsion and a coupling with phononic degrees of freedom. In the standard second quantization formalism, the Hamiltonian reads
\begin{equation}
    \begin{split}
        \mathcal{H} = & -t\sum_{ \ave{\textbf{ij}}, \sigma } \big( c^\dagger_{\mathbf{i}\sigma} c_{\mathbf{j}\sigma} + \text{H.c.} \big) -\mu \sum_{\textbf{i},\s} n_{\textbf{i}\s} + \sum_\textbf{i} U_\textbf{i} n_{\textbf{i}\uparrow} n_{\textbf{i}\downarrow}\\
        & + \sum_\textbf{i} \bigg( \dfrac{\hat{P}_\textbf{i}^2}{2M}+  \dfrac{ M\omega_0^2 \hat{X}_\textbf{i}^2 }{2} \bigg) - g \sum_{\textbf{i},\s} n_{\textbf{i}\s} \hat{X}_\textbf{i}
    \end{split}
    \label{HHH}
\end{equation}
where $c^\dagger_{\mathbf{i}\sigma} (c_{\mathbf{i}\sigma} )$ are creation (annihilation) operators of electrons with spin $\sigma$ at a given site $\textbf{i}$, while $n_{\textbf{i}\s} \equiv c^\dagger_{\mathbf{i}\sigma} c_{\mathbf{i}\sigma} $ are number operators.
Here, the sums run over a two-dimensional square lattice under periodic boundary conditions, with $\ave{\textbf{ij}}$ denoting nearest-neighbor sites.
The first two terms on the right hand side of Eq.\,\eqref{HHH} correspond to the kinetic energy of electrons, and their chemical potential $\mu$ term, respectively, while the on-site Coulomb repulsion between electrons is included through the third term.
Notice that we have introduced a site dependence on the interaction strength, $U_{\mathbf{i}}$, which is described in detail below. 
The phonon degrees of freedom appear in the fourth term as quantum harmonic oscillators with frequency $\omega_0$ (as an Einstein model), with $\hat{P}_\textbf{i}$ and $\hat{X}_\textbf{i}$ being conjugate momentum and position operators, respectively. 
The last term corresponds to the electron-ion coupling, whose strength is $g$.
Hereafter, we define the mass of the ions, $M$, and the lattice and Boltzmann constants as unity, while using the hopping integral, $t$, to define the scale of energy. 

At this point, it is important to recall that the electron-phonon coupling leads to \textit{polarons}, i.e.~quasiparticles formed by electrons dressed by a cloud of phonons, whose characteristic energy scale is $\lambda = g^2 / \omega_0^2$.
Therefore, it is convenient to adopt $\lambda / t$ as the strength of the $e$-$ph$ interaction.
In addition, we also define $\omega_0 / t $ as the adiabaticity ratio.
Throughout this work, we have fixed $\lambda /t = 2$, and $\omega_0 / t =1$, while varying the Coulomb strength, $U/t$, and the fraction \textit{x} of magnetic impurities.

We investigate the properties of Eq.\,\eqref{HHH} by performing finite temperature determinant quantum Monte Carlo (DQMC) simulations \cite{Blankenbecler81,Hirsch83,Hirsch85,scalettar89}. The DQMC approach is an unbiased method based on the decoupling of the non-commuting terms of the Hamiltonian in the partition function by Trotter-Suzuki decomposition, i.e., by discretizing the inverse of temperature into small imaginary-time steps,  $\beta = M \Delta\tau$.
For the Hubbard-Holstein Hamiltonian, such a procedure leads to
\begin{align}
\nonumber
\mathcal{Z}&=\mathrm{Tr}\,e^{-\beta\widehat{\mathcal{H}}} \\
& \approx
\mathrm{Tr}\,
[\cdots e^{-\Delta\tau\widehat{\mathcal{H}}_{K}} e^{-\Delta\tau\widehat{\mathcal{H}}_{ph}} e^{-\Delta\tau\widehat{\mathcal{H}}_{\rm
el-ph}} e^{-\Delta\tau\widehat{\mathcal{H}}_{\rm
U}} \cdots],
\end{align}
with error of $\mathcal{O}(\Delta\tau^2)$, but becoming exact for $\Delta\tau \to 0$.
Here, $\mathcal{H}_{K}$, $\mathcal{H}_{\rm ph}$, $\mathcal{H}_{\rm el-ph}$, and $\mathcal{H}_{U}$ correspond to the kinetic, bare phonon modes, electron-phonon coupling, and Hubbard interaction terms, respectively.

To proceed, we employ a Hubbard-Stratonovich transformation to obtain the  $e^{-\Delta\tau\widehat{\mathcal{H}}_{\rm U}}$ operators as quadratic forms, but with the price of adding new degrees of freedom $s_{{\bf i},l}$: the Hubbard-Stratonovich fields.
The bosonic and fermionic traces `Tr' lead to
\begin{align}\label{eq:partition_func_Hols}
{\cal Z}  \propto \int \mathcal{D}\{x_{{\bf i},l}\}& \mathcal{D}\{s_{{\bf i},l}\}  \,e^{-\Delta \tau S_{B}} \times 
\nonumber \\
& \Pi_{\sigma} \bigg[ \mathrm{det}\big(I + B^{\sigma}_{M} B^{\sigma}_{M-1} \cdots B^{\sigma}_{1} \big) \bigg],
\end{align}
with $x_{{\bf i},l}$ being phonon degrees of freedom, and $S_{B}$ the bare phonon action,
\begin{align}\label{eq:phonon_action}
S_{B} = \frac{\omega_{0}^2}{2} \sum_{{\bf i}}
 \sum_{l=1}^{M} \bigg[ 
\frac{1}{\omega_{0}^2 \Delta\tau^2} \big( x_{{\bf i},l} - x_{{\bf i},l+1} \big)^2 
+  x^{2}_{{\bf i},l}  \bigg].
\end{align}
The matrices $B^{\sigma}_{l} \equiv B^{\sigma}_{l}(x_{{\bf i},l},s_{{\bf i},l})$ result from a product of an exponential of the kinetic term and site-diagonal matrices with the exponential of the electron-phonon and Hubbard terms, at a given imaginary time slice $l$.
The integrals $\int \mathcal{D}\{x_{{\bf i},l}\} \mathcal{D}\{s_{{\bf i},l}\}$ -- i.e., the bosonic traces -- are performed by Monte Carlo methods.
Within this approach, one may obtain both equal-time and unequal-time Green's functions, $\mathcal{G^{\sigma}(\tau,\tau^\prime)}$, and, therefore, any higher-order correlation functions.
Further methodological details may be found, e.g., in Refs.\,\cite{Kawashima2002,rrds2003,sorella2017}.

\begin{figure}[t]
    \centering
    \includegraphics[scale=0.5]{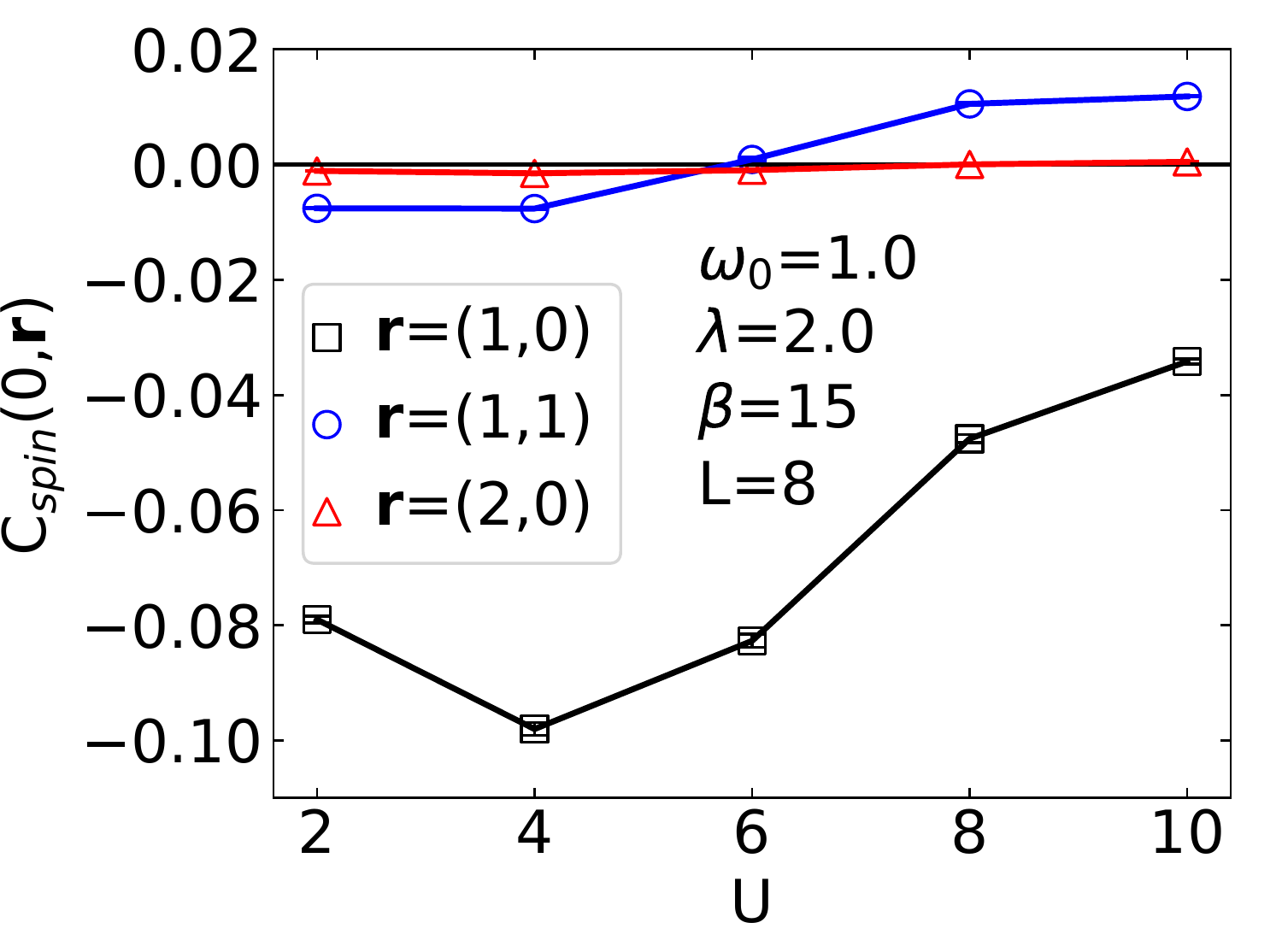}
    \caption{Spin-spin correlation functions between the impurity site and the three nearest neighbors, as a function of $U$.
    }
    \label{sosj1imp} 
\end{figure}

Given this, we investigate the magnetic properties of the Hamiltonian in Eq.\,\eqref{HHH} through the spin-spin correlation functions, 
\begin{equation}
	C_\text{spin}(\textbf{i},\textbf{j} )=   \langle  (n_{\textbf{i}\uparrow} - n_{\textbf{i}\downarrow} )(n_{\textbf{j}\uparrow} - n_{\textbf{j}\downarrow} ) \rangle, 
\end{equation} 
while its charge response is examined through the charge-charge ones, 
\begin{equation}
	C_\text{charge}(\textbf{i},\textbf{j} ) = \langle(n_{\textbf{i}\uparrow} + n_{\textbf{i}\downarrow} )(n_{\textbf{j}\uparrow} + n_{\textbf{j}\downarrow} ) \rangle.
\end{equation}
In particular, we probe the occurrence of charge instabilities through the behavior of the charge structure factor, 
\begin{equation}
S_{\text{cdw}}(\textbf{q}) = \frac{1}{N} \sum_{\textbf{i}, \textbf{j}} e^{-i\textbf{q} \cdot (\textbf{i}- \textbf{j})} C_\text{charge}(\textbf{i},\textbf{j} ),
\end{equation}
and its correlation ratio
\begin{equation}
    R_{\text{cdw}}(L) = 1 - \dfrac{ S_{\text{cdw}} (\textbf{Q} +\delta \textbf{q})}{S_{\text{cdw}}(\textbf{Q})}
    \label{eq:corratio}
\end{equation}
with $N = L \times L$ being the number of sites, $ | \delta \textbf{q}| = 2\pi/L$, and $\textbf{Q} = (\pi, \pi)$.
The crossing of $R_{\rm cdw}(L)$ for different lattice sizes, together with their finite-size scaling analyses, provides estimates for the location of the critical region \cite{Kaul2015,Gazit2018,sato2018,liuhong2018,darmawan2018}.

Finally, we also examine the pairing response by means of the finite temperature pair susceptibility
\begin{align}
\chi_\text{sc}(\alpha) = \frac{1}{N} \sum_{\textbf{i,j}} \int_0^\beta  \langle \Delta_\alpha (\textbf{i},\tau) \Delta^\dagger_\alpha (\textbf{j},0) \rangle d\tau~,
\end{align}
with 
\begin{equation}
	\Delta _\alpha (\textbf{i},\tau) = \frac{1}{2} \sum_\textbf{a} f_\alpha (\textbf{a}) c_{\textbf{i}\downarrow}(\tau) c_{\textbf{i+a}\uparrow} (\tau), 
\end{equation}
where $c_{\textbf{i}\s}(\tau)= e^{\tau \mathcal{H}} c_{\textbf{i}\s} e^{-\tau \mathcal{H}}$, and $f_\alpha (\textbf{a})$ is the pairing form factor for a given wave symmetry $\alpha=s$ or $d$.
Under some circumstances, such as in the present case, it is more adequate to remove the noninteracting (vertex) contribution to the susceptibility,  $\bar{\chi}_\text{sc}(\alpha)$, and define the effective response as
$\chi_\text{sc}^\text{eff}(\alpha) = \chi_\text{sc}(\alpha) - \bar{\chi}_\text{sc}(\alpha)$; see, e.g., Ref.\,\cite{White89}.

In order to describe the effects of magnetic impurities in a CDW environment, here we deal with the half-filled square lattice, which is known to exhibit a charge-ordered ground state for any finite electron-phonon coupling, $\lambda/t > 0$.
We model the impurities by allowing for random distributions of $U_{\mathbf{i}} > 0$ on the lattice, for a given fraction $x$ of the sites, as
\begin{equation}
U_{\mathbf{i}}= \left \{
    \begin{array}{c l}	
         U & {\rm with\,\,probability}\,\,x; \\
         0 & {\rm with\,\,probability}\,\,(1-x) .
    \end{array}\right.
\end{equation}
To ease the discussion, in what follows we present the results for the dilute regime (i.e.\ one and two impurities) separate from those for the dense regime of many impurities. All results for the dense limit were obtained by averaging the quantities over 100-200 disorder configurations. 

\begin{figure}[t]
    \centering
    \includegraphics[scale=0.5]{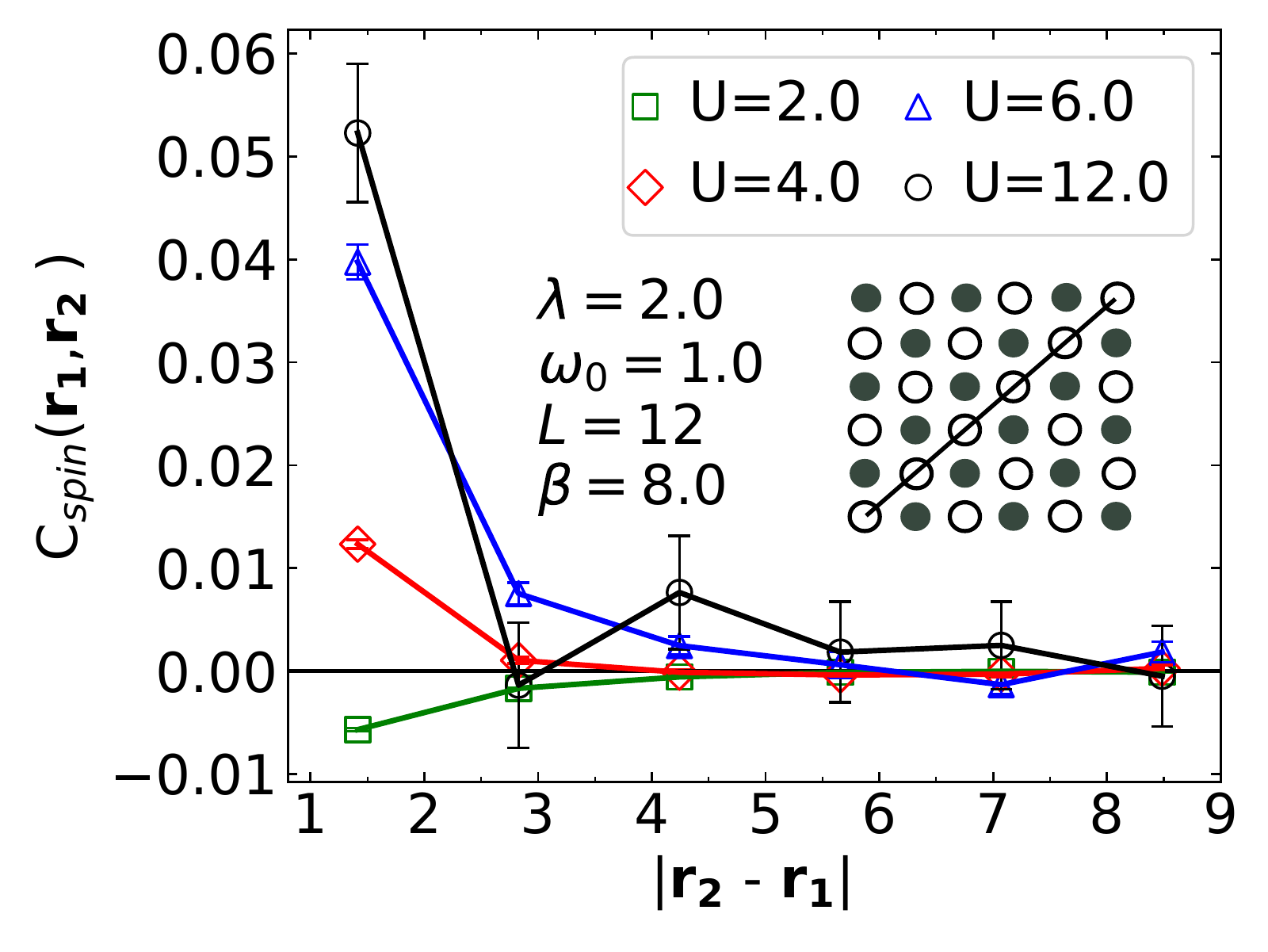}
    \caption{Spin-spin correlation functions between two impurities on the same sublattice, as a function of the distance $|\textbf{r}_2 - \textbf{r}_1|$ between them, and for different values of $U/t$.
    }
    \label{s1s2r}
\end{figure}

\section{Results for the dilute case}
\label{sec:dilute}


\begin{figure*}
    \centering
    \includegraphics[scale=0.95]{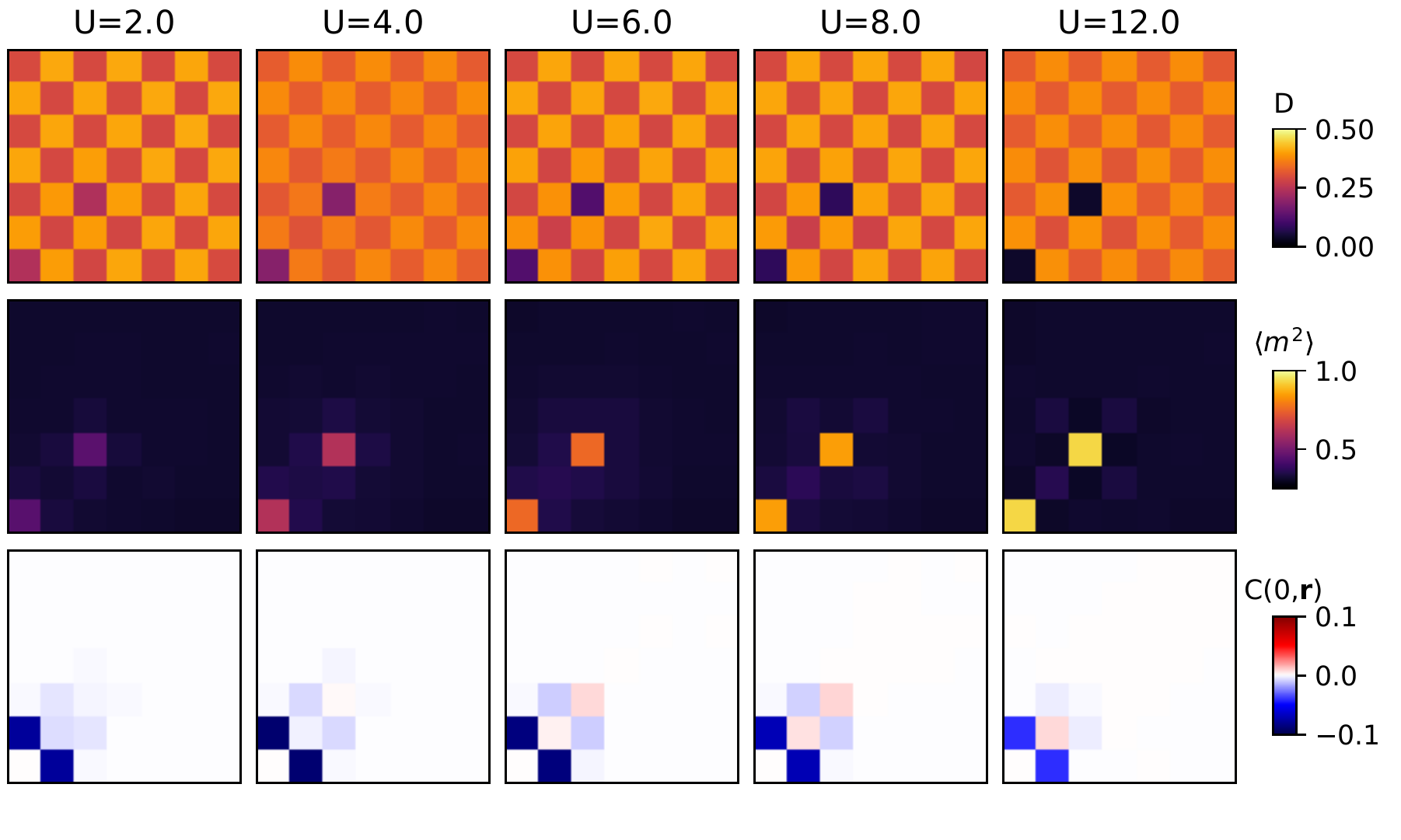}
    \caption{Real space results for the double occupancies (upper panels), local moments (middle panels), and charge-charge correlation functions (lower panels) for a system with two impurities, one in the origin, and the other the coordinates $(2,2)$. Here we fixed $L=12$.}     
    \label{fig:profiles}
\end{figure*}

\subsection{Single impurity}
\label{subsec:single}
We start with the investigation of the single-impurity case. 
For the sake of comparison, we recall that when a magnetic impurity is placed in a \emph{metallic} environment ($\lambda =0$), AFM correlations are enhanced around it, whose strength decays with distance; 
in addition, as the number of impurities increases, this AFM cloud evolves towards a long-range ordered configuration\,\cite{Ulmke98b,Lima20}.
In a CDW background, on the other hand, such staggered spin-spin correlations are drastically suppressed or even destroyed, as displayed in Figure \ref{sosj1imp}.
Notice that for small values of $U$, first-, second-, and third-neighbor spin-spin correlations exhibit negative responses, in stark contrast with the previous picture of an AFM cloud in the metallic case.
In order to understand this difference, note that the effects brought about by a single $U$-impurity in a charge-ordered background start with the broken two-fold degeneracy of the CDW ground state on a square lattice: if the impurity is located on, say the $\alpha$-sublattice, the CDW is stabilized on the $\beta$-sublattice.
The very weak spin correlations between the impurity site and its second- and third-neighbors is accounted for by the fact these sites belong to the $\alpha$-sublattice, hence with a very small local moment when $U$ is small.
More robust spin correlations with sites on the $\beta$-sublattice, on the other hand, indicate a redistribution of the spin cloud surrounding the impurity.
As $U$ increases, nearest neighbor correlations are first strengthened and then weakened\,\footnote{This weakening of the nearest neighbor spin correlation functions is due to thermal effects.}, which is accompanied by a reversal of the sign of $c(1,1)$ (blue circles in Fig.\,\ref{sosj1imp}): this indicates a new redistribution such 
that antiferromagnetic correlations become dominant locally, with the creation of an AFM cloud around the impurity.
That is, in the presence of a CDW background, a large $U$ is needed to generate an AFM cloud around the impurity; 
such ``critical'' value of $U$ has a strong dependence with the electron-phonon coupling strength (not shown).
The occurrence of this AFM cloud is what leads to the emergence of long-range order for the many-impurity case, discussed below.

\subsection{Two impurities}
\label{subsec:double}

Let us now consider the case of two impurities, and examine how the relative position between the $U_\iv\neq0$ centers affects the overall properties.
Figure \ref{s1s2r} shows the spin-spin correlation functions between two impurity sites on the same sublattice as a function of their distance.
Similarly to the single-impurity case, the profile of correlations is very sensitive to the magnitude of $U$.
For $U=~2$ and when the impurity sites are nearest neighbors within the same sublattice (NN$\alpha$), the impurity spins are weakly antiferromagnetically correlated, while for $U=4$ they are already ferromagnetically correlated.
As $U$ is increased further, the profile changes considerably, in the sense that the period of oscillation seems to decrease. 

The first row of Figure \ref{fig:profiles} shows the profile of double occupancy, 
\begin{equation}
	D_\iv\equiv \ave{n_{\iv\up} n_{\iv\dn}},
\label{eq:dblocc}	
\end{equation}
with $D_\iv\in [0,1/2]$ at half filling, for different values of $U$; the impurity sites are placed on the same sublattice, at $(x,y)=(0,0)$ and  $(x,y)=(2,2)$. 
As expected, the sublattice symmetry is broken, with $D_\iv$ being always larger on the sublattice not containing the impurities. 
Further, as $U$ increases, $D_\iv$ on the impurity sites decreases steadily.
The middle row of Figure \ref{fig:profiles} shows the local moment, 
\begin{equation}
	\ave{m_\iv^2}\equiv \ave{(n_{\iv\up} -n_{\iv\dn})^2}
		=[\ave{n_\iv}- 2D_\iv],
\end{equation}	
with $\frac{1}{N}\sum_\iv\ave{n_\iv}=n=1$ at half filling, for different values of $U$.  
It is clear that the local moment is suppressed on all sites, except on the impurity ones.
We may thus conclude that there is an increasing tendency to occupy the impurity sites with a single spin.

The bottom row of Figure \ref{fig:profiles} shows the spin correlations between an impurity site (placed at the origin) and sites with coordinates $\mathbf{r}\equiv(x,y)$, in the presence of the second impurity at $(x,y)=(2,2)$.
If $U$ is small, the sites surrounding the impurity tend to align antiferromagnetically with it, which includes, although less intense, the other impurity; in view of the analysis of the single impurity case, the presence of a second impurity strengthens the AFM correlations around the impurities. 
However, for increasing values of $U$, the correlations along the diagonal display oscillations with $U$.
Further, the period of oscillation depends on the relative position between the sites, reminiscent of an RKKY-like interaction, but having in mind that this occurs in the presence of a CDW background. 



\begin{figure}[t]
    \centering
    \includegraphics[scale=0.5]{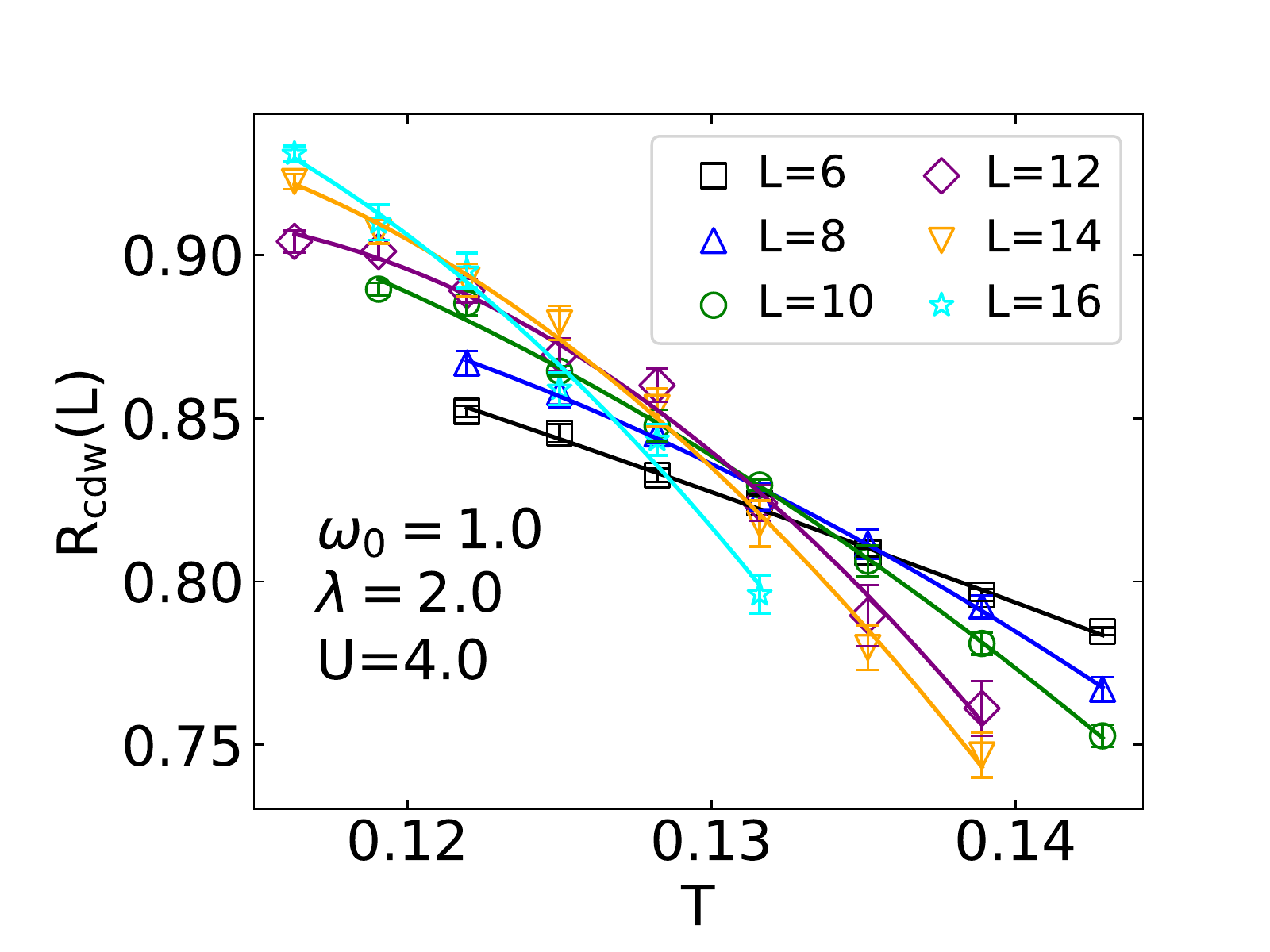}
    \caption{The CDW correlation ratio [Eq.\,\eqref{eq:corratio}] for fixed $\omega_0$, $\lambda$, and $U$ as a function of temperature, for $x = 0.2$ and different lattice sizes, $L$. The crossings provide estimates for $T_c$.}
    \label{rcdw}
\end{figure}

\section{The dense regime}
\label{sec:dense}
\subsection{CDW transition}
\label{subsec:cdwdense}

In the previous section we established that AFM correlations are able to overcome the charge order locally by increasing $U$ in a single impurity, or in two repulsive sites. 
In this section we aim to determine the minimum concentration of $U$-sites required to destroy the CDW for $U > \lambda$. 
Since the presence of impurity sites tends to deplete doubly occupied sites, a decrease in the CDW critical temperature is expected to occur for increasing $x= N_\text{imp}/N$, where $N_\text{imp}$ is the number of impurity sites with $U\neq 0$.


\begin{figure}[t]
    \centering
    \includegraphics[scale=0.5]{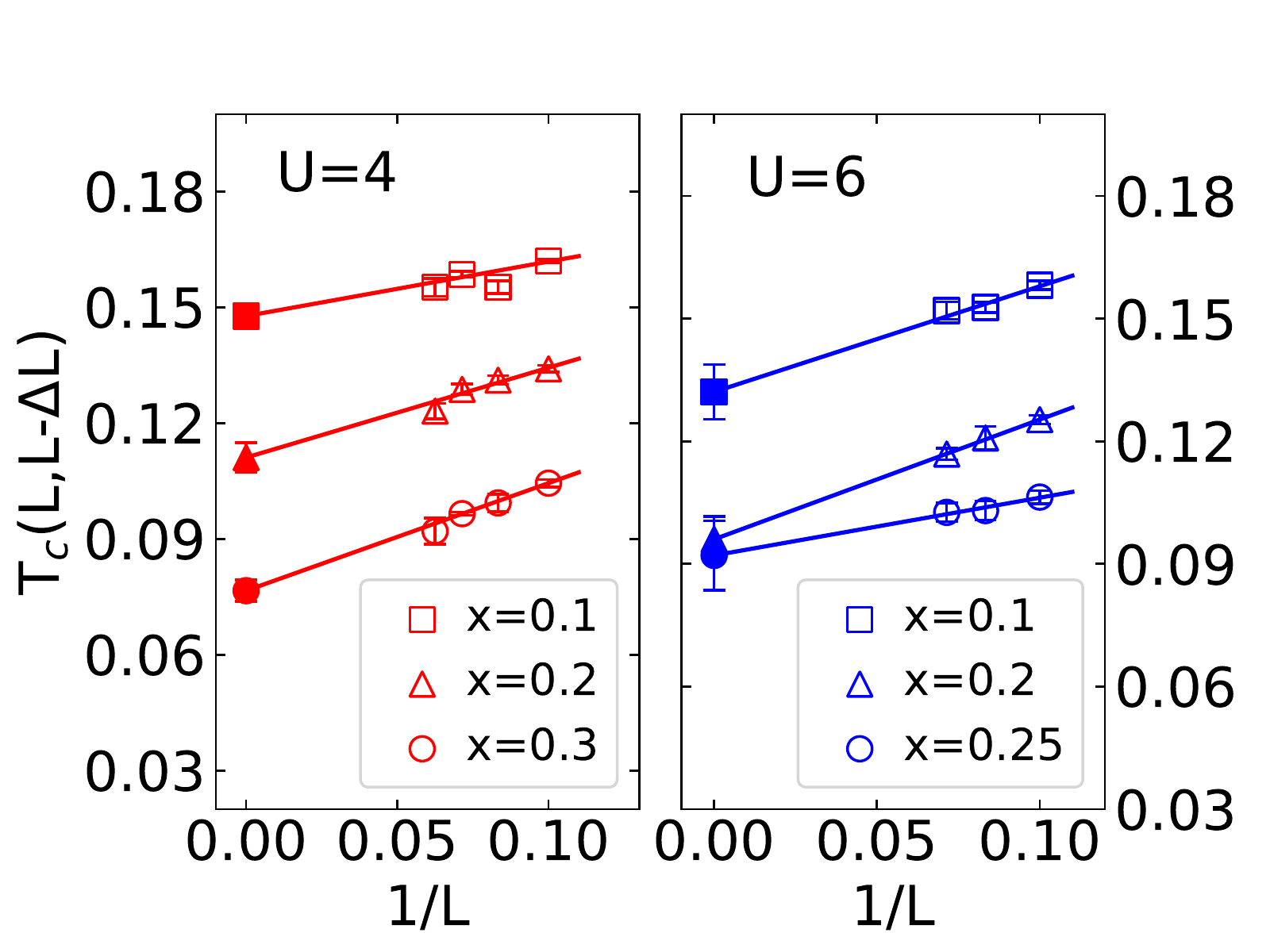}
    \caption{Finite-size scaling plots of the estimated CDW critical temperatures as determined from intersects such as those in Fig.\,\ref{rcdw}, using $\Delta L=4$; see text. Empty symbols are the estimates, and filled symbols are the extrapolated values.}
    \label{tcl}
\end{figure}

\begin{figure}[t]
    \centering
    \includegraphics[scale=0.53]{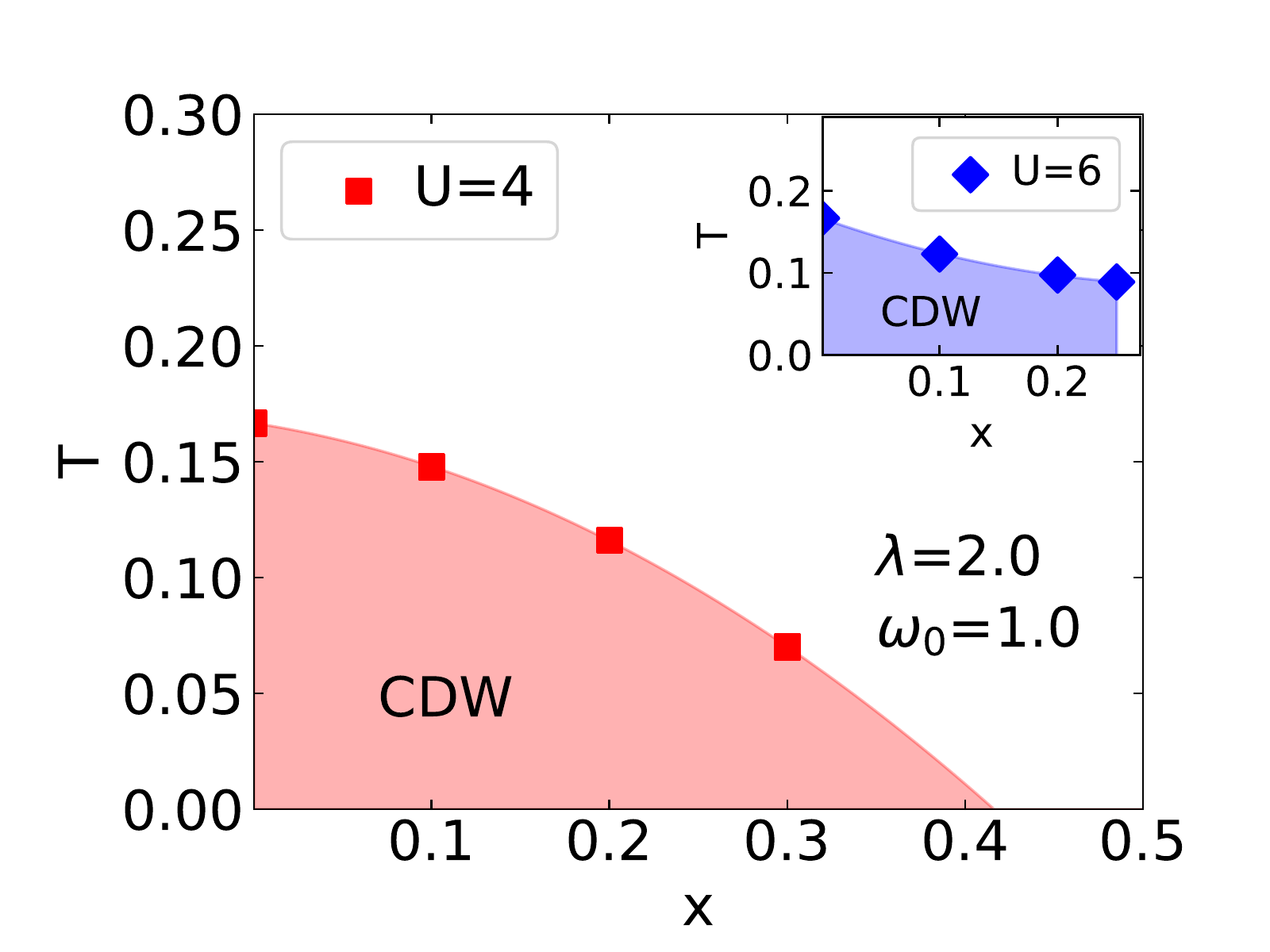}
    \caption{The CDW critical temperature as a function of impurity concentration, for $U/t=4$, and the same parameters of Fig.\,\ref{tcl}. Inset: same analysis, but for $U/t=6$. The error bars are smaller than the data points, and the lines are guides to the eye. 
    }
    \label{tcxc}
\end{figure}

For the many-impurity case, we recall that the quantities of interest are obtained after performing configurational averages. 
Figure \ref{rcdw} shows the configurationally averaged  CDW correlation ratio $R_{\text{cdw}}(L)$ [Eq.\,\eqref{eq:corratio}] as a function of temperature, for different system sizes. 
The temperatures at which two curves, $R_{\rm cdw}(L)$ and $R_{\rm cdw}(L-\Delta L)$, intersect provides an estimate $T_c(L, L-\Delta L)$. 
Figure \ref{tcl} exhibits these crossings for $U=4$ and $U=6$, and for fixed $\Delta L = 4$, from which we may extrapolate towards the critical temperature in the thermodynamic limit (see, e.g., the filled symbols).
By extrapolating these estimates for $1/L\to0$, we obtain the phase diagrams $T_c$ versus $x$ displayed in Fig.\,\ref{tcxc}, from which we see that $T_c$ decreases with increasing $x$, as expected.
Further, the data in Fig.\,\ref{tcxc} provides an estimate for the critical disorder concentration above which there is no CDW at finite temperatures for $U=4$, namely $x_c\approx 0.41$.
The same estimation for $U=6$ is challenging due to the minus sign problem, which is more severe at low temperatures, and larger values of $U$. 

Nonetheless, it is interesting to note that the $U\neq 0$ sites play the role of disordering agents as far as CDW order is concerned, so that $x$ corresponds to the concentration $1-p$ of inactive sites in ordinary percolation \cite{Stauffer03}; given that the critical site percolation threshold for the square lattice is $p_c=0.59$ (see, e.g.\ Ref.\,\cite{Stauffer03}), our estimate $x_c=0.41$ may indicate a major role played by geometrical constraints. 
This should be contrasted with a recent study of the Hubbard model on a disordered Lieb lattice, which shows that the concentration threshold for magnetism is strongly dependent on the on-site repulsion \cite{Lima20}. 
For the present case, it is not a coincidence that our $x_c$ is close to the geometric percolation threshold of the square lattice.
One may understand this from the results of the single- and two-impurity cases.
As discussed in Fig.\,\ref{sosj1imp}, an AFM cloud is not formed when $U/t=4$ and $\lambda/t = 2$, requiring that impurities should be sited side-by-side in order to have strong spin-spin correlations.
Indeed, this picture is confirmed in the two impurities case, where Figs.\,\ref{s1s2r} and \ref{fig:profiles} show a drastic suppression of the spin-spin correlation functions for $U/t \lesssim 8$.
That is, due to these short-range correlations, one may expect the CDW phase to be destroyed close to the classical geometric percolation threshold, even at moderately large values of $U/\lambda$. Therefore, $x_c$ should be unchanged for $U/t=6$, although the minus-sign problem prevents us from presenting numerical data.

\color{black}

\subsection{The insulating state}
\label{subsec:insulator}

In a clean system ($x=0$), the temperature-driven CDW transition leads to an insulating state at half-filling, characterized by a Peierls gap and absence of superconductivity\,\cite{scalettar89,Weber18,ncc2018}.  
We will now discuss how the presence of repulsive centers affects the transition to the insulating state. 
To this end we resort to several quantities such as the double occupancy, $D$ [Eq.\,\eqref{eq:dblocc}],  the compressibility,
\begin{equation}
	\kappa = \frac{\beta}{N} \sum_{\iv,\jv} \langle \delta n_\iv \delta n_\jv \rangle, 
\end{equation}
where 
\begin{equation}
	\delta n_\iv\equiv \sum_\sigma [n_{\iv\sigma} - \ave{n_{\iv\sigma}}], 
\end{equation}	
the kinetic energy $E_k$, and the uniform static spin susceptibility, 
\begin{equation}
	\chi_\text{sp} = \frac{\beta}{N} \sum_{\textbf{i},\,\textbf{j}}C_\text{spin}(\textbf{i},\textbf{j}). 
\end{equation}
Further analyses of these quantities may provide signatures of the crossover into a bad-metallic phase, as discussed in Ref.\,\cite{Aaram2020}. 

\begin{figure}[t]
    \centering
    \includegraphics[scale=0.53]{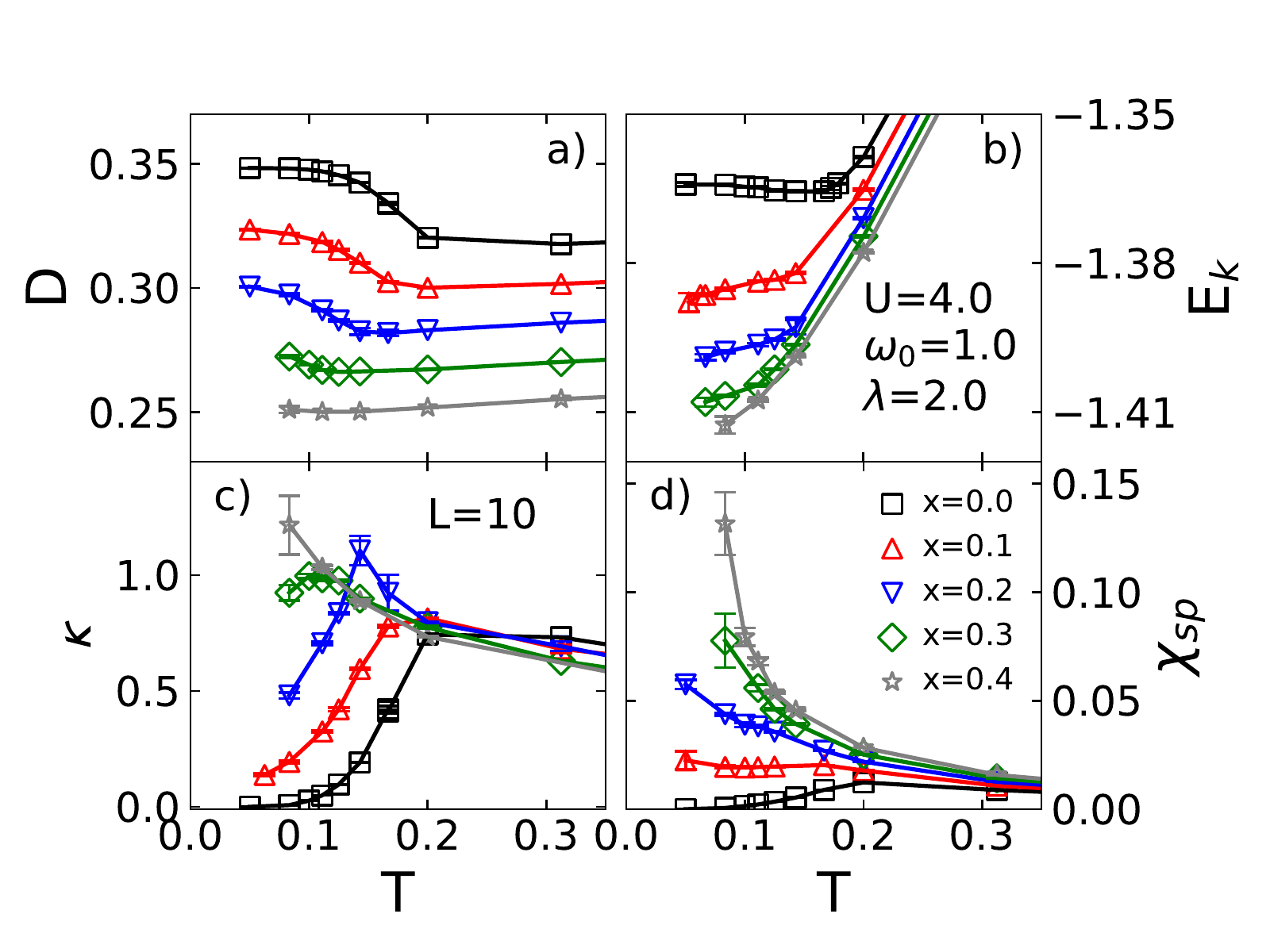}
    \caption{(a) Double occupancy, (b) kinetic energy, (c) compressibility, and (d) uniform static spin susceptibility as functions of temperature, $T/t$. Each curve is for a given impurity concentration, and all data are for fixed values of $\omega_0$, $\lambda$, and $U$.}
    \label{obsT}
\end{figure}

Figure \ref{obsT}\,(a) displays the behavior of the double occupancy as a function of temperature.
For a CDW ground state, one expects a large value of $D$ below $T\lesssim T_c$, while for a noninteracting metallic state one has $D=0.25$.
Given this, notice that a peak in $\partial D(x,T) / \partial T$ appears for temperatures close to critical ones (see, e.g., Fig.\,\ref{tcxc}), with the exception of $x=0.4$, which presents $D \approx 0.25$ at low-$T$.
This is in agreement with our previous analysis, for the destruction of the CDW ground state for $x_c \approx 0.4$.
Similar observations apply to the compressibility [Fig.\,\ref{obsT}\,(c)], for which the change in slope is accompanied by a peak, whose positions decrease with increasing disorder.

However, the behavior of the kinetic energy, exhibited Fig.\,\ref{obsT}\,(b), is more subtle.
The black square symbols show its behavior for the clean case, where the ground state has a well-formed Peierls gap.
For such a case, notice that the kinetic energy exhibits a slight increase for $T \lesssim T_c$, clearly showing an insulating behavior.
In the presence of disorder, however, the kinetic energy still decreases within the insulating phase, although at a smaller rate than in the metallic phase.
That is, we still have an insulating phase, but a \textit{bad insulator}, when compared to the Peierls one.
We shall return to this discussion below.

Finally, Fig.\,\ref{obsT}\,(d) shows the temperature dependence of the uniform spin susceptibility for different impurity concentrations. 
For the clean system, the susceptibility goes to zero exponentially, reflecting the presence of a spin gap due to the doubly occupied sites forming the CDW state. 
As discussed before, the impurities tend to form antiferromagnetic clouds around them by depleting the doubly occupied sites. 
For several impurities, however, these clouds display no long range antiferromagnetic order, so that a uniform magnetic field may easily polarize the local moments at the impurity sites, and a Curie-like magnetic response sets in as $x$ increases, explaining the disappearance of the spin gap.

\subsection{Superconducting and spectral properties}
\label{subsec:gaps}

\begin{figure}[t]
    \centering
    \includegraphics[scale=0.53]{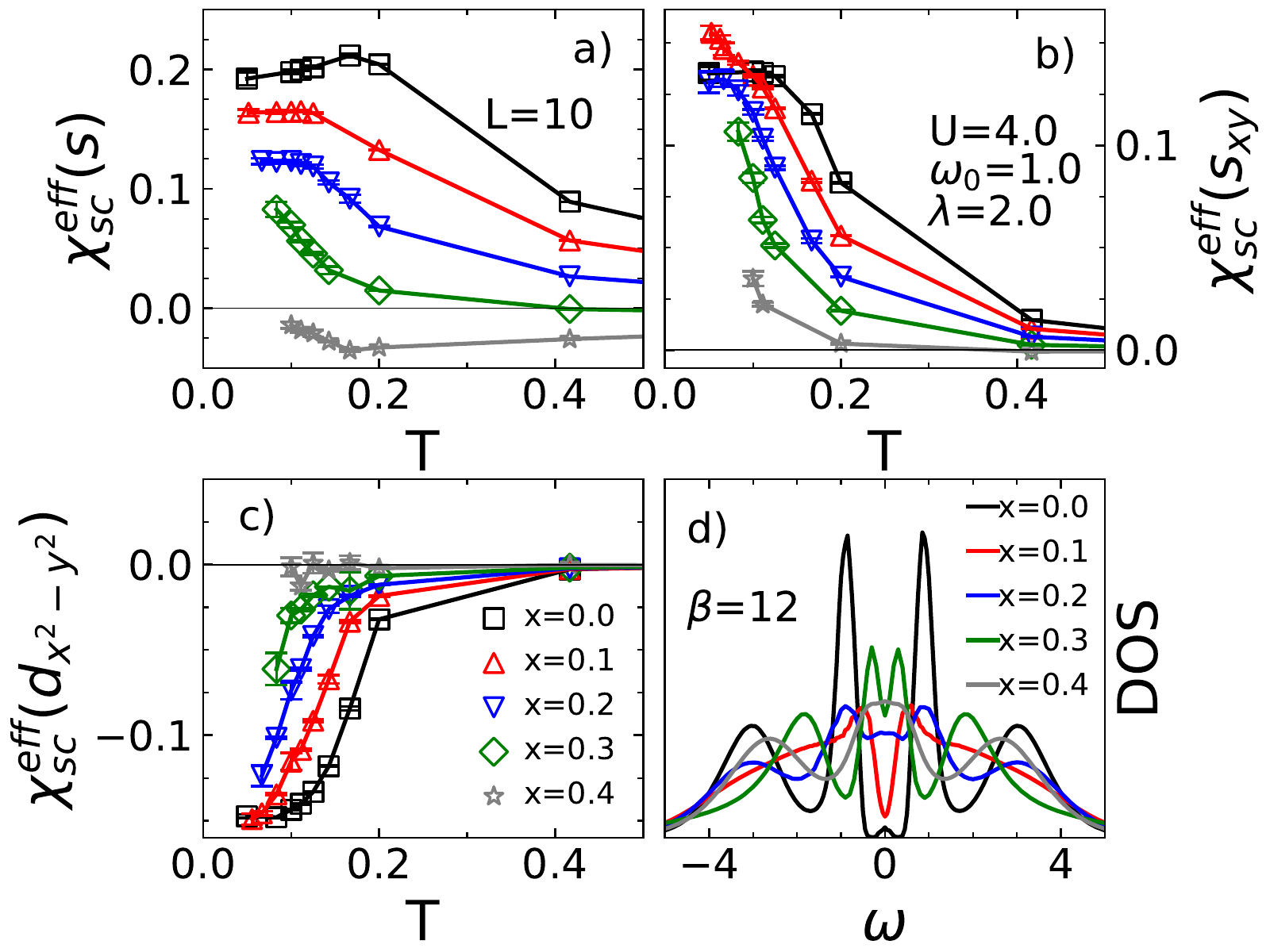}
    \caption{Effective pairing susceptibility $\chi_\text{sc}^\text{eff}$ as a function of $T$ for: (a) $s$-wave, (b) $s_{xy}$-wave,  and (c) $d_{x^2-y^2}$-wave channels. 
    Panel (d) shows the density of states (DOS) for $\beta/t = 12$ and several impurity concentrations. In each panel, the curves are for the given impurity concentrations, and all data are for fixed values of $\omega_0$, $\lambda$, and $U$.}
    \label{Peff}
\end{figure}

Having established that magnetic impurities destroy the CDW insulating state, one may wonder whether this can favor superconducting correlations.
In order to check this, we have calculated the effective pairing susceptibility $\chi_\text{sc}^\text{eff}(\alpha)$ for $s$-wave, $s_{xy}$-wave, and $d_{x^2-y^2}$-wave symmetries; when this quantity is positive, an attractive channel sets in.
Figure \ref{Peff}(a)-(c) shows the temperature dependence of the effective pairing susceptibility, from which we may rule out any pairing tendencies in the $d_{x^2-y^2}$ symmetry [panel (c)] -- notice that $\chi_\text{sc}^\text{eff}$ is strongly suppressed as the temperature is lowered for all $x$.
For the $s$- and $s_{xy}$-wave channels, on the other hand, the dirty system displays a behavior different from the clean one: $\chi_\text{sc}^\text{eff}$ increases as $T$ decreases.

Although the pairing tendency is enhanced with the presence of impurities in a CDW background, pairing correlations are unable to drive the system to a long-range superconducting ordered state in the regime of electron-phonon coupling considered here; a much larger $\lambda$ is required to unequivocally reach long-range order.
This is evidenced by the stabilization of $\chi_\text{sc}^\text{eff}$ at low temperatures, around the critical CDW temperature -- long-range order would require a divergence in $\chi_\text{sc}^\text{eff}$, which does not seem to occur for the range of temperatures analyzed.
By contrast, in the absence of a CDW background, for $x > x_{c}$, $\chi_\text{sc}^\text{eff} > 0$ may be interpreted as favoring the formation of a superconducting phase.
Unfortunately, for $x \gtrsim x_{c}$, the average fermionic sign is small, thus preventing analyses at low temperatures.
At any rate, it is interesting to notice that for $x = 0.4$ the only attractive pairing channel is the $s_{xy}$-wave.
That is, a superconducting state emerging in this region would display a pairing symmetry different from the standard on-site $s$-wave, due to the Coulomb electron-electron interaction, in agreement with Refs.\,\cite{ncc20,Wang20}.

Finally, it is also worth examining the spectral properties of the system.
We compute the density of states (DOS) by performing an analytic continuation of the imaginary-time dependent Green's function, using the Maximum
Entropy Method\,\cite{Jarrell96};
this amounts to inverting the integral equation
\begin{align}
\mathcal{G}(\mathbf{r}_{ij}=0,\tau) = \int \mathrm{d}\omega\,
N(\omega) \,
\frac{e^{-\omega \tau}}{e^{\beta \omega} + 1},
\label{eq:aw}
\end{align}
with $\mathbf{r}_{ij}$ denoting the relative
displacement between sites, and $N(\omega)$ is the sought DOS.
Figure \ref{Peff}\,(d) shows the evolution of the DOS as the impurity concentration increases, for fixed $\beta/t = 12$.
We see that disorder suppresses the Peierls gap present in the clean system, even for a small amount of impurities, e.g., for $x=0.10$.
Although it seems contradictory with the compressibility results in Fig.\,\ref{obsT}\,(c), which shows an insulating state for these values of $x$ and $T$, it gives support for the results of Fig.\,\ref{obsT}\,(d), which suggests the appearance of localized states.
Therefore, the scenario brought about by these results is that magnetic impurities suppress the Peierls gap by creating in-gap localized states, which, in turn,  enhance magnetic correlations while destroying the CDW background at the percolation threshold. 

\section{Conclusions}
\label{sec:concl}

We have studied the effect of magnetic impurities interacting with a charge density wave background, stabilized by electron-phonon coupling within the Holstein model scenario. 
The impurities are modelled by assigning a repulsive Hubbard-$U$ coupling to a site, which tends to favor the formation of a local moment, and we have considered a square lattice with a half-filled electronic band.

By first analyzing the dilute regime (one and two impurities) we have established that, unlike the metallic case, only for large $U$ the local moment is significant and an AFM cloud forms around the isolated impurity site; this results from depleting nearby doubly occupied sites. 
When two impurities are placed on the same sublattice, spin correlations between them oscillate with $U$, reminiscent of an RKKY behavior.
However, these correlations are strongly suppressed with distance, so that only nearest or next-nearest neighbors are relevant.
In the dense regime, the impurities lower the critical temperature for CDW formation, which vanishes at some critical impurity concentration, $x_c$.
Interestingly, our data for $U=4$ yield $x_c \approx 0.4$, consistent with the classical percolation threshold for the square lattice (0.41).
That is, due to the short-range charge-charge correlations the destruction of the CDW phase depends on geometrical aspects.

However, the CDW state that emerges when magnetic impurities are present is not a regular Peierls one.
We have also established that the Peierls and the spin gaps are both suppressed by even a small amount of impurities.
The occurrence of such a \textit{bad insulating} phase is due to localized states at the Fermi level filling the gap, whose local moments, in turn, give rise to a Curie-like magnetic response.
For $x \lesssim x_c$, these magnetic impurities are not able to suppress the CDW background, but may drastically change thermodynamic properties, such as the average kinetic energy. 

Our data also show that superconducting correlations are enhanced in the $s$- and $s_{xy}$ channels, as a result of suppression of the CDW state.
This is consistent with recent experimental findings \cite{Dai1993} of an increase in the superconducting temperature with intercalation of Fe in CDW materials.
For $x \gtrsim x_c$, when magnetic impurities destroy the CDW background, we expect that long-range order should emerge, but for nonlocal (not on-site) pairs, as an effect of the electron-electron interaction.
In closing, we note that it has been recently suggested~\cite{Xiao2021} that Anderson-like disorder in the Holstein model also gives rise to an enhancement of superconducting correlations at low temperatures, in agreement with our overall findings that disorder significantly disturbs the CDW state, and leads to pairing.

\color{black}

\section*{ACKNOWLEDGMENTS}
Financial support from the Brazilian Agencies CAPES, CNPq, and FAPERJ, and Instituto Nacional de Ci\^encia e Tecnologia de Informa\c c\~ao Qu\^antica (INCT-IQ) is gratefully acknowledged.
N.C.C.~and S.A.~Sousa-J\'unior acknowledge PRACE for awarding them access to Marconi at CINECA, Italy.
N.C.C.~acknowledges financial support from CNPq, grant number 313065/2021-7.

\bibliography{ref.bib}

\end{document}